\documentclass[a4paper,onecolumn,sloppy,12pt]{scrartcl}
\usepackage{setspace}
\usepackage[T1]{fontenc}
\usepackage{lmodern}
\usepackage[super]{natbib} 
\usepackage{longtable}
\usepackage{microtype}
\usepackage{bbm}
\usepackage[footnotesize,labelfont=bf]{caption}
\DeclareCaptionLabelSeparator{bar}{ | }
\usepackage{graphicx}
\usepackage{subfigure}
\usepackage{color}
\usepackage{amsmath}

\usepackage{hyperref}
\usepackage{authblk}
\usepackage{parskip}
\usepackage{fancyhdr}
\setkomafont{sectioning}{\rmfamily\bfseries\boldmath}

\newcommand{\We}{W\hspace{-0.25em}e}

\title{Pancake bouncing on superhydrophobic surfaces}

\author[1]{\small{Yahua Liu}\footnote{These authors contributed equally to the paper.}}
\author[2]{\small{Lisa Moevius}$^\ast$}
\author[3]{\small{Xinpeng Xu}}
\author[3]{\small Tiezheng Qian}
\author[2]{\small Julia M Yeomans\footnote{Corresponding author, j.yeomans1@physics.ox.ac.uk}}
\author[1,4]{\small Zuankai Wang\footnote{Corresponding author, zuanwang@cityu.edu.hk}}
\affil[1]{\small Department of Mechanical and Biomedical Engineering, City University of Hong Kong, Hong Kong}
\affil[2]{\small The Rudolf Peierls Centre for Theoretical Physics, 1 Keble Road, Oxford, OX1 3NP, UK}
\affil[3]{\small Department of Mathematics, Hong Kong University of Science and Technology, Clear Water Bay, Kowloon, Hong Kong}
\affil[4]{\small Shenzhen Research Institute of City University of Hong Kong, Shenzhen, China}

\date{}

\begin{document}

\maketitle

\begin{abstract}
	Engineering surfaces that promote rapid drop detachment\cite{QuereRichard2002, VaranasiBird2013} is of importance to a wide range of applications including anti-icing\cite{PoulikakosJung2012, AizenbergMishchenko2010, Stone2012}, dropwise condensation\cite{WangChen2011}, and self-cleaning\cite{Blossey2003, CohenTuteja2007, ButtDeng2012}. Here we show how superhydrophobic surfaces patterned with lattices of submillimetre-scale posts decorated with nano-textures can generate a counter-intuitive bouncing regime: drops spread on impact and then leave the surface in a flattened, pancake shape without retracting. This allows for a four-fold reduction in contact time compared to conventional complete rebound \cite{QuereRichard2002, QuereOkumura2003, QuereReyssat2006, Bartolo2006, WangMcCarthy2012}.
	We demonstrate that the pancake bouncing results from the rectification of capillary energy stored in the penetrated liquid into upward motion adequate to lift the drop. Moreover, the timescales for lateral drop spreading over the surface and for vertical motion must be comparable.
In particular, by designing surfaces with tapered micro/nanotextures which behave as  harmonic springs, 
 the timescales become independent of the impact velocity, allowing the occurrence of pancake bouncing and rapid drop detachment over a wide range of impact velocities.
\end{abstract}
\vspace{1cm}

\begin{figure}[t!] 
  \centering
  \includegraphics[width=1\textwidth]{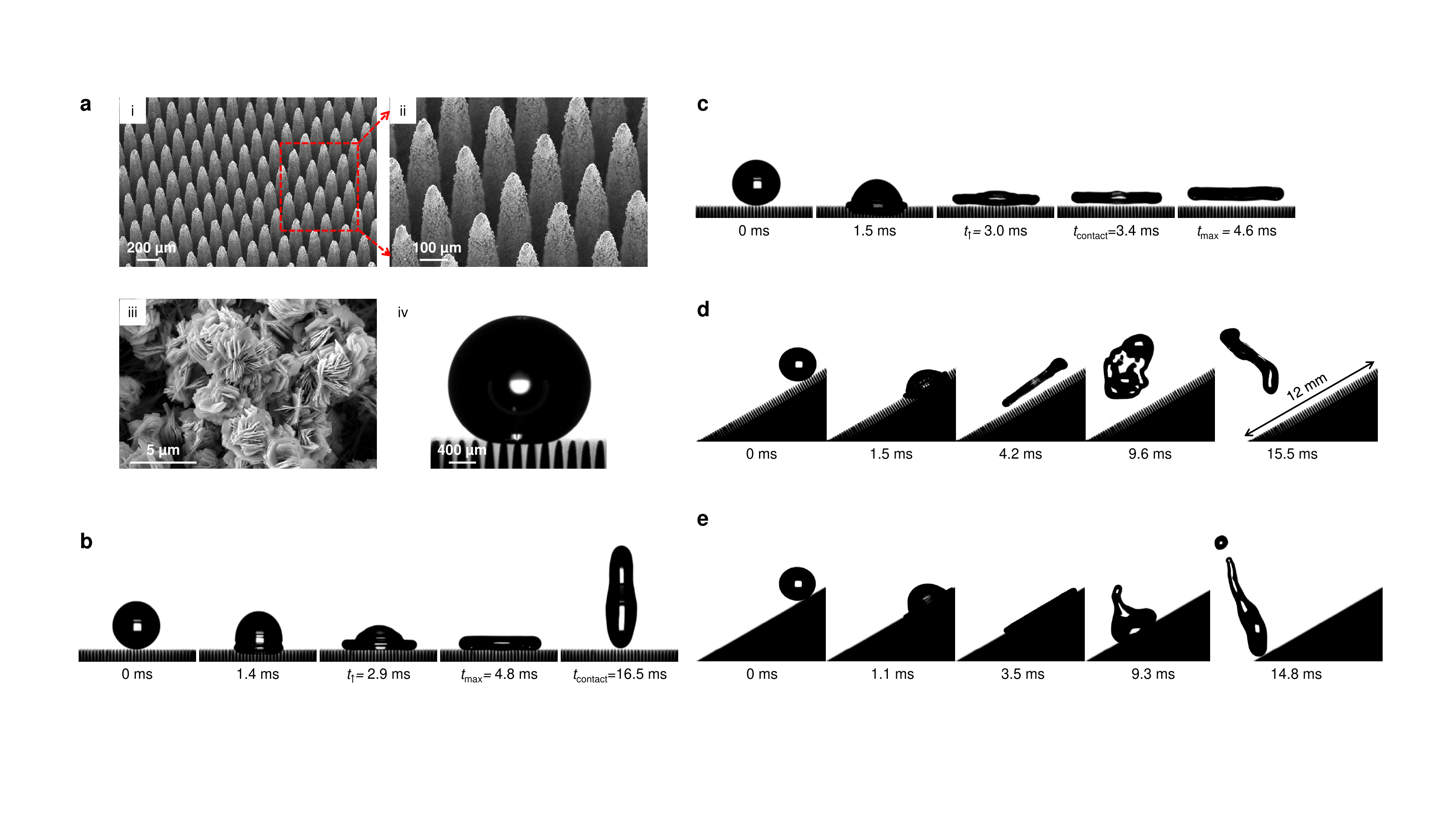} \\
  \caption{ \textbf{Surface characterization and drop impact dynamics.} \textbf{a}, Scanning electronic micrograph image of the copper surface patterned with a square lattice of tapered posts. The posts have a circular cross section whose diameter increases continuously and linearly from $20 \, \mathrm{\mu m}$ to $90 \, \mathrm{\mu m}$ with depth. The center-to-center spacing and the height of the post are $200 \, \mathrm{\mu m}$ and $800 \, \mathrm{\mu m}$, respectively. The posts are covered by nanoflowers of average diameter $3.0 \, \mathrm{\mu m}$, exhibiting a contact angle of over $165^{\circ}$ and contact angle hysteresis less than $2^{\circ}$.
\textbf{b}, Selected snapshots captured by the high speed camera showing a drop ($r_0 = 1.45\, \mathrm{mm}$) impacting on the tapered surface at $\We=7.1$. Upon touching the surface at $t=0$, part of the drop penetrates into the post arrays and recoils back (driven by capillary force) to the top of the surface at $t_{\uparrow} \sim$ 2.9\,ms (Supplementary Movie 1\cite{SI}). After reaching a maximum lateral extension at $t_{max} \sim 4.8$\,ms, the drop retracts on the surface and finally detaches from the surface at $t_{contact}$ ($\sim 16.5$\,ms). \textbf{c}, Selected snapshots showing a drop impacting on the tapered surface at $\We=14.1$. The drop bounces off the surface in a pancake shape at $\sim$ 3.4\,ms.  (\textbf{d}, \textbf{e}), Selected snapshots showing a drop impinging on the tapered surface and superhydrophobic surface with nanoflower structure alone, respectively, under a tilt angle of $30^{\circ}$ at $\We=31.2$. The drop impinging on the tapered surface exhibits a pancake bouncing (\textbf{d}), while the drop on the nanostructured surface follows a conventional bouncing pathway (\textbf{e}). The contact time in the case of pancake bouncing is 3.6\,{ms}, which is four-fold shorter than that on the nanostructured superhydrophobic surface.}
  \label{fig:fig1}
\end{figure}

Consider a copper surface patterned with a square lattice of tapered posts decorated with nanostructures\cite{BartoloMoulinet2007, BhushanJung2008, KimCha2010, Tran2013} (Fig.~\ref{fig:fig1}a). The post height $h$ is $800 \, \mathrm{\mu m}$ and the centre-to-centre spacing $w$ is $200 \, \mathrm{\mu m}$ (Supplementary Fig.~1a\cite{SI}). The posts have a circular cross section with a diameter which increases continuously and linearly from $20 \, \mathrm{\mu m}$ to $90 \, \mathrm{\mu m}$ with depth in the vertical direction.  The post surface is fabricated using a wire cutting machine followed by chemical etching\cite{BhushanJung2008,ZhangChen2009,Tran2013} to generate nanoflowers of average diameter $3.0 \, \mathrm{\mu m}$. After a thin polymer coating, trichloro(1H,1H,2H,2H-perfluorooctyl)silane, is applied, the surface exhibits a superhydrophobic property with an apparent contact angle of over $165^{\circ}$ (Fig.~\ref{fig:fig1}a). The advancing and receding contact angles are $167.2^{\circ}\pm 1.1^{\circ}$ and $163.9^{\circ} \pm 1.4^{\circ}$, respectively.
Water drop impact experiments were conducted using a high speed camera at the rate of 10,000 frames per second. The unperturbed radius of the drop is $r_0=1.45 \, \mathrm{mm}$ or $1.10 \,\mathrm{mm}$, and the impact velocity ($v_0$) ranges from $0.59 \, \mathrm{m \, s^{-1}}$ to $1.72 \, \mathrm{m \, s^{-1}}$, corresponding to $7.1 < {\We} < 58.5$, where $\We=\rho v_0^2 r_0/\gamma$  is the Weber number, with $\rho$ the density and $\gamma$ the surface tension of water.
\par
Fig.~\ref{fig:fig1}b shows selected snapshots of a drop impinging on such a surface at ${\We}=7.1$. Upon touching the surface at $t=0$, part of the drop penetrates into the post arrays in a localized region with the radius approximately equivalent to the initial drop radius and recoils back, driven by the capillary force, to the top of the surface at 2.9\,ms (Supplementary Movie\cite{SI}~1). After reaching a maximum lateral extension\cite{QuereClanet2004} at 4.8\,ms, the drop retracts on the surface and finally detaches from the surface at 16.5\,ms ($=2.55\sqrt{\rho r_0^3/\gamma}$). This contact time is in good agreement with previous results for conventional complete rebound \cite{QuereRichard2002,QuereOkumura2003, QuereReyssat2006,Bartolo2006}. However, at higher $\We$, the drop exhibits a distinctively different bouncing behaviour, which we term pancake bouncing, as exemplified by an impact at $\We=14.1$ (Fig.~\ref{fig:fig1}c, Supplementary Movie\cite{SI}~2). In this case, the liquid penetration is deeper and the drop detaches from the surface (at 3.4\,ms $=0.53\sqrt{\rho r_0^3/\gamma}$) immediately after the capillary emptying without experiencing retraction.
\par

\begin{figure}[t!] 
  \centering
\includegraphics[width=1.0\textwidth]{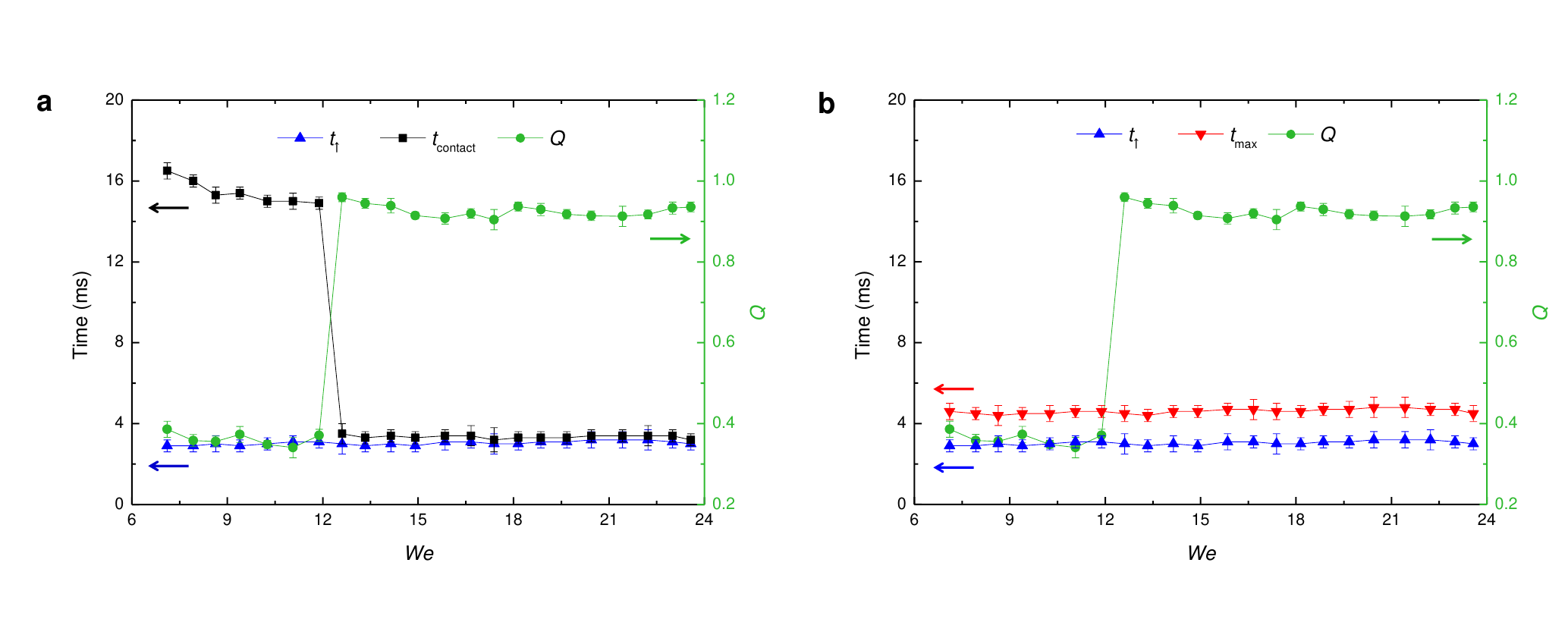} \\
  \caption{ \textbf{Timescale analysis of drop impact on tapered surface.} \textbf{a}, The variations of $t_{\uparrow}$, $t_{contact}$ (left $y$ axis), and pancake quality $Q$ ($= d_{jump}/d_{max}$, right $y$ axis) with $\We$ for drop radius $r_0 = 1.45\, \mathrm{mm}$. At low $\We<12$, the drop exhibits conventional bouncing with $t_{contact}$ much larger than $t_{\uparrow}$. However, at high $\We>12$ the drop bounces as a pancake with $t_{\uparrow} \approx t_{contact}$. \textbf{b}, The variations of $t_{\uparrow}$, $t_{max}$ (left $y$ axis) and $Q$ (right $y$ axis) with $\We$. $t_{\uparrow}$ and $t_{max}$ are nearly constant over a wide range of $\We$ from $8$ to $24$. Each data point is the average of three measurements. Error bars denote the range of the measurements.
}
  \label{fig:fig2}
\end{figure}

The difference in bouncing dynamics between conventional rebound and pancake bouncing can be quantified by the ratio of the diameter of the drop when it detaches from the surface $d_{jump}$ to the maximum spreading width of the drop $d_{max}$. The ratio $Q= d_{jump}/ d_{max}$ is defined as the pancake quality, with $Q>0.8$ referred to as pancake bouncing. At low Weber number ($\We<12.6$), the pancake quality $Q$ is $\sim 0.4$, corresponding to conventional bouncing\cite{QuereRichard2002,QuereOkumura2003,QuereReyssat2006,Bartolo2006,WangMcCarthy2012} (Fig.~\ref{fig:fig2}a). However, for $\We>12.6$ there is a clear crossover to  $Q\sim 1$, which corresponds to pancake bouncing. Moreover, a defining feature of pancake bouncing, of particular relevance to applications aimed at rapid drop shedding, is the short contact time\cite{QuereRichard2002,VaranasiBird2013} of the drop with the solid surface. In the case of pancake bouncing, the contact time, $t_{contact}$, is reduced by a factor of over four to 3.4\,{ms} as compared to conventional rebound\cite{QuereRichard2002,QuereOkumura2003,QuereReyssat2006,Bartolo2006,WangMcCarthy2012}.
\par
We also performed drop impact experiments on tilted surfaces, a geometry more relevant to practical applications, such as self-cleaning\cite{Blossey2003,CohenTuteja2007,ButtDeng2012}, de-icing \cite{PoulikakosJung2012, AizenbergMishchenko2010,Stone2012} and thermal management\cite{WangChen2011, Vakarelski2012}. Fig.~\ref{fig:fig1}d shows selected snapshots of a drop impinging on the tapered surface with a tilt angle of $30^{\circ}$ at ${\We}=31.2$ (Supplementary Movie\cite{SI}~3, left). The drop impinging on the tilted tapered surface also exhibits pancake bouncing. Moreover, the drop completely detaches from the surface within 3.6\,{ms} and leaves the field of view without bouncing again. We also compared the drop impact on the tilted surface with nanoflower structure alone. The apparent contact angle of the nanostructured surface is $160^{\circ}\pm 1.8^{\circ}$. It is evident that drop impinging on such a surface follows a conventional bouncing pathway: the drop spreads to a maximum diameter, recoils back, and finally leaves the surface within 14.5\,ms (Fig.~\ref{fig:fig1}e, Supplementary Movie\cite{SI}~3, right).

\par
We propose that the pancake bouncing of a drop occurring close to its maximum lateral extension results from the rectification of the capillary energy stored in the penetrated liquid\cite{BartoloMoulinet2007,YarinLembach2010, ButtDeng2013} into upward motion adequate to lift the entire drop.
Moreover, for the drop to leave the surface in a pancake shape, the timescale for the vertical motion between posts should be comparable to that for the lateral spreading.
To validate that pancake bouncing is driven by the upward motion rendered by the capillary emptying, we compared the two timescales $t_{contact}$ and $t_{\uparrow}$, where $t_{\uparrow}$ is the time interval between the moment when the drop first touches the surface and when the substrate is completely emptied, during which fluid undergoes the downward penetration and upward capillary emptying processes (Supplementary Fig.\cite{SI}~2). As shown in Fig.~\ref{fig:fig2}a, in the regime of pancake bouncing, $t_{contact}$ and $t_{\uparrow}$ are close, indicating that the pancake bouncing is driven by the upward motion of the penetrated liquid\cite{YarinLembach2010,ButtDeng2013}. For smaller $\We$ ($< 12.6$), the two time scales diverge: $t_{\uparrow}$ remains approximately constant while $t_{contact}$ increases sharply. This is because, at low $\We$, the penetrated liquid does not have the kinetic energy sufficient to lift the drop at the end of the capillary emptying. Accordingly, the drop continues to spread and retract in contact with the surface before undergoing conventional bouncing  \cite{QuereRichard2002,QuereOkumura2003,QuereReyssat2006,Bartolo2006,WangMcCarthy2012}.
Next, we plotted the variations of $t_{\uparrow}$, $t_{max}$, and $Q$ with $\We$ (Fig.~\ref{fig:fig2}b), where $t_{max}$ is the time when the drop reaches its maximum lateral extension\cite{QuereOkumura2003,QuereClanet2004}. On tapered surfaces, $t_{\uparrow}$ and $t_{max}$ are comparable with each other for all the $\We$ measured. However, at low $\We$ ($<12.6$), there is no pancake bouncing due to insufficient energy to lift the drop, further indicating that the occurrence of pancake bouncing necessitates the simultaneous satisfaction of sufficient impact energy and comparable timescales.

\begin{figure}[t!] 
  \centering
  \includegraphics[width=1\textwidth]{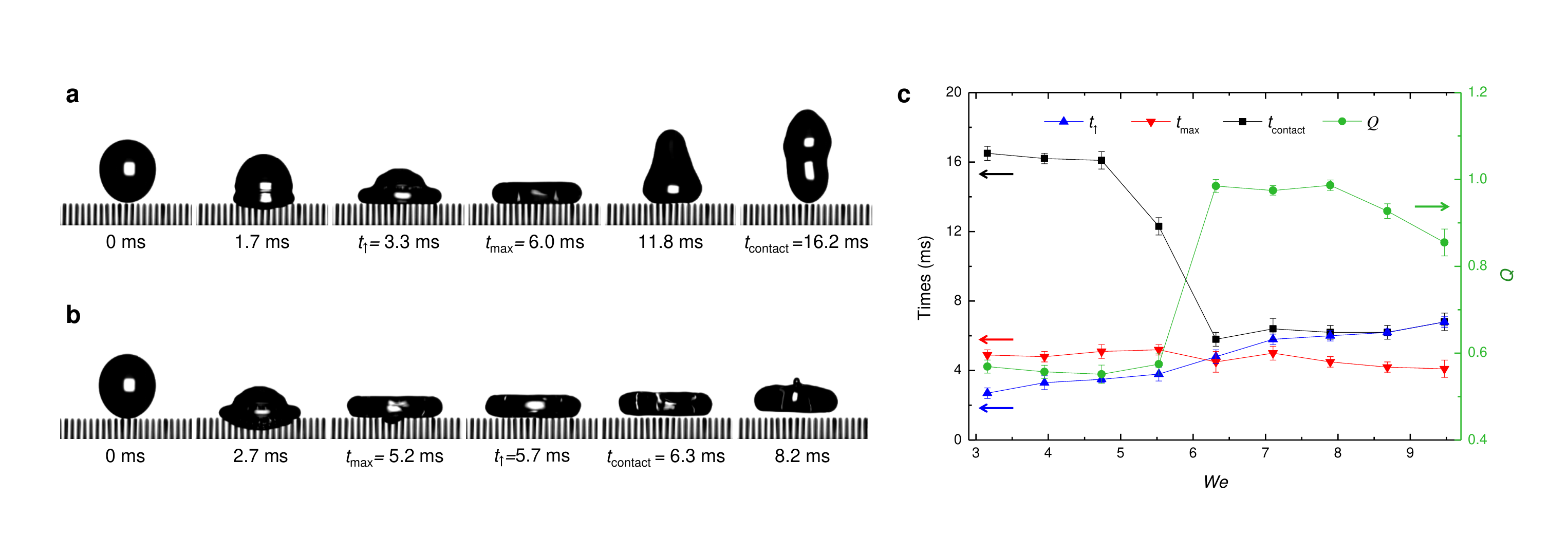}\\
  \caption {\textbf{Drop impact dynamics on straight square posts decorated with nanoflowers.} \textbf{a}, Selected snapshots of a drop impinging on straight posts decorated with nanoflowers with a post centre-to-centre spacing of $300 \, \mathrm{\mu m}$ at $\We=4.7$. The drop exhibits conventional rebound with pancake quality $Q \sim 0.59$.  $t_{max} \sim$ 6.0\,ms is much larger than $t_{\uparrow} \sim$ 3.3\,ms. \textbf{b}, Selected snapshots of a drop impinging on post arrays with a post centre-to-centre post spacing of $300 \, \mathrm{\mu m}$ at $\We=7.9$. Pancake bouncing is observed with pancake quality $Q \sim 0.98$, $t_{max} \sim $ 5.2\,ms is slightly less than $t_{\uparrow} \sim$ 5.7\,ms. \textbf{c}, The variations of $t_{\uparrow}$, $t_{max}$, $t_{contact}$ (left $y$ axis), and $Q$ (right $y$ axis) with $\We$. $t_{max}$ is a constant over a range of $\We$ from $3$ to $10$. At low $\We<6.3$, the drop exhibits conventional bouncing with $t_{contact}$ much larger than $t_{\uparrow}$. However, at high $\We>6.3$, the drop bounces in the shape of a pancake with $t_{\uparrow} \approx t_{contact}$. Unlike on the tapered surfaces, $t_{\uparrow}$ increases with increasing $\We$.}
  \label{fig:fig3}
\end{figure}
\par
We next compared experimental results for bouncing on straight square posts covered by nanoflower structures. The post height and edge length ($b$) are $1.2 \, \mathrm{mm}$ and $100 \, \mathrm{\mu m}$, respectively (Supplementary Fig.\cite{SI}~1b). We observed that the pancake bouncing behavior is sensitive to post spacing and $\We$.
Pancake bouncing is absent on post arrays with $w=200 \, \mathrm{\mu m}$ (Supplementary Fig.\cite{SI}~3), whereas it occurs for surfaces with $w=300 \, \mathrm{\mu m}$ and $400 \, \mathrm{\mu m}$.
Fig.~\ref{fig:fig3}a and b compare results for the bouncing of a drop ($r_0 \sim 1.45 \, \mathrm{mm}$) on the surface with spacing $300 \, \mathrm{\mu m}$ at $\We=4.7$ and $7.9$, respectively.
In the former case, the drop exhibits a conventional complete rebound, with $Q \sim 0.59$ and $t_{contact} \sim $ 16.2\,ms.
In the latter case, the drop shows pancake bouncing with $Q \sim 0.98$ and a much reduced contact time $t_{contact} \sim$ 6.3\,ms (Supplementary Movie\cite{SI}~4).
Fig.~\ref{fig:fig3}c shows the variations of $t_{\uparrow}$, $t_{max}$, $t_{contact}$, and $Q$ with $\We$ for this surface.
In the region of pancake bouncing ($6.3\le \We \le 9.5$), the proximity of $t_{contact}$ and $t_{\uparrow}$ and the matching between $t_{max}$ and $t_{\uparrow}$ are consistent with the observations on tapered surfaces. By contrast, in the non-pancake bouncing region ($\We \le 6.3$), there is a large divergence between $t_{contact}$ and $t_{\uparrow}$, because $\We$ is too small to allow drop bouncing as a pancake.
This further confirms that the occurrence of pancake bouncing necessitates simultaneous satisfaction of the two criteria.
Different to tapered surfaces, a dependence of $t_{\uparrow}$ on $\We$ is noted to appear on straight posts.
Moreover, we found the maximum jumping height of drops in pancake shape on straight posts is three-fold smaller than that on tapered surfaces ($2.88 \, \mathrm {mm}$ and $0.9 \, \mathrm{mm}$, respectively) and that the contact time ($\sim$ 6.3\,ms) on straight posts is larger than that ($\sim$ 3.4\,ms) on tapered surfaces. All these observations reveal that the pancake bouncing on tapered surfaces is more pronounced and robust than that on straight posts.
\par
We propose a simple analytical argument to elucidate the enhanced pancake bouncing observed on tapered posts in comparison to straight posts.
The timescale $t_{max}$ scales as $\sqrt{{\rho}r_0^3/{\gamma}}$, independent of the impact velocity \cite{QuereRichard2002,QuereOkumura2003,QuereClanet2004,Bartolo2006,QuereReyssat2006}.
To calculate $t_{\uparrow}$, we consider the kinetics involved in the processes of liquid penetration and capillary emptying. Here, we neglect the viscous dissipation\cite{NagelXu2005} since the Reynolds number in the impact process is $\sim 100$.
The liquid penetrating into the space between posts is subject to a capillary force, which serves to halt and then reverse the flow. The capillary force can be approximated by $ b n \gamma \cos{\theta_Y}$ \cite{Bartolo2006, QuereReyssat2006,BartoloMoulinet2007}, where $n$ is the number of posts wetted, and $\theta_Y$ is the intrinsic contact angle of the nanoflower-covered posts. The deceleration (acceleration) of the penetrated liquid moving between the posts scales as  $a_\uparrow  \sim {b  \gamma  \cos{\theta_Y }}/({\rho r_0 w^2})$, where the drop mass $\sim \rho r_0^3$, $n \sim { r_0^2}/{w^2}$, and we assume that the liquid does not touch the base of the surface. Note that the number of posts wetted is independent of $\We$ because the penetrating liquid is mainly localized in a region with a lateral extension approximatively equivalent to the initial drop diameter, rather than the maximum spreading diameter (Supplementary Figs.\cite{SI}~4,~5). For straight posts, the acceleration is constant. Thus, $t_\uparrow \sim v_0/a_\uparrow  \sim  {v_0 \rho  r_0 w^2}/{(- b  \gamma  \cos{\theta_Y })}$, and the ratio of the two timescales can be expressed as
\begin{equation}\label{scaling}
k ={ t_\uparrow}/{t_{max}} \sim \sqrt{\We} \frac{w^2}{(-b r_0\cos{\theta_Y})},
\end{equation}
which scales as $\sqrt{\We}$. Our experimental observations show, as discussed previously, that the occurrence of pancake bouncing requires $t_\uparrow$ and ${t_{max}}$ to be comparable, i.e., $k\sim 1$. The dependence of $k$ on $\We$ indicates that this condition can be satisfied only over a limited range of $\We$.
\par
Interestingly, $k$ and $\We$ become decoupled by designing surfaces with tapered posts.
Since the post diameter $b$ now increases linearly with the depth $z$ below the surface (i.e., $b \sim \beta z$, where $\beta$ is a structural parameter), the acceleration of the penetrated liquid moving between posts is linearly proportional to penetration depth (i.e., $a_{\uparrow} \propto z$).  As a result, the surface with tapered posts acts as a harmonic spring with $t_\uparrow \sim \sqrt{{w^2 r_0 \rho}/{(-\beta \gamma  \cos{\theta_Y } )}} $. 
Therefore, the ratio of timescales becomes
\begin{equation}
 k \sim \frac{w}{r_0 \sqrt{- \beta \cos{\theta_Y }} },
\end{equation}
which is independent of $\We$.

\begin{figure}[t!] 
  \centering
  \includegraphics[width=1\textwidth]{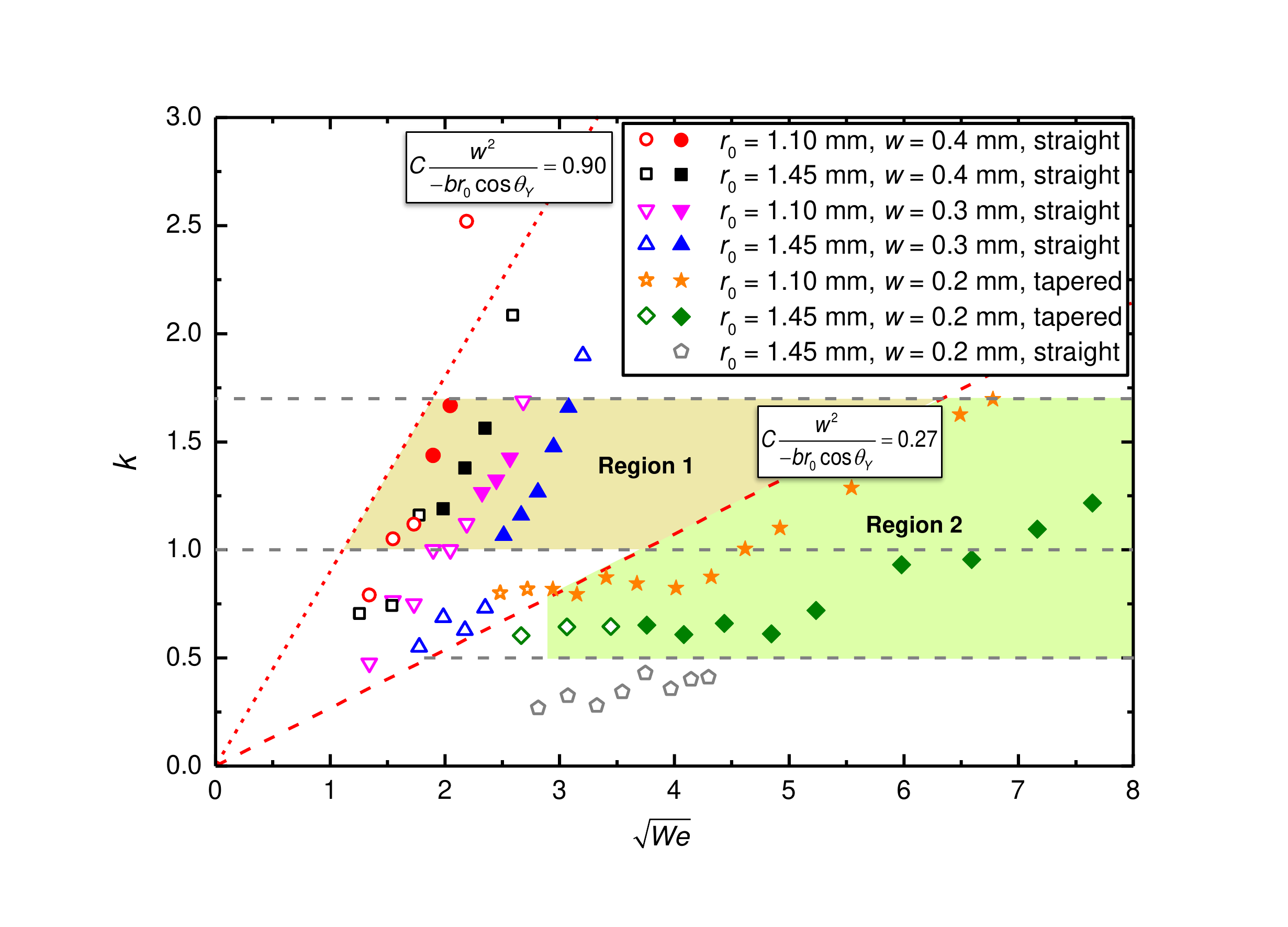}\\
  \caption {\textbf{Design diagram.} The variation of the timescale ratio $k=t_{\uparrow}/t_{max}$ with $\sqrt{\We}$, showing different pancake bouncing regions. Full symbols denote that the drop jumps as a pancake. Region 1 corresponds to the pancake bouncing on straight posts with $1.0<k<1.7$ and $\We$ in a restricted range. The two slanting lines, corresponds to $-w^2/b r_0\cos{\theta_Y}= 0.45$ and $1.5$ (based on Eq.~(1), with a fitting prefactor $C=0.6$). Region 2 corresponds to  pancake bouncing on tapered surfaces over a much wider range of $k$ from 0.5 to 1.7 and $\We$ from 8.0 to 58.5. Note that $k$ is independent of $\We$ over a wide range. It becomes weakly dependent on $\We$ for higher impact velocities due to the penetrated liquid hitting the base of the surface.}
  \label{fig:fig4}
\end{figure}
\par
To pin down the key surface features and drop parameters for the occurrence of pancake bouncing, we plotted the variation of $k$ with $\sqrt{\We}$ in the design diagram (Fig.~\ref{fig:fig4}). Solid symbols represent pancake bouncing (defined by $Q>0.8$) and open symbols denote conventional bouncing. Region 1 corresponds to the pancake bouncing occurring on straight posts with $1.0<k<1.7$. The data show that $k\sim\sqrt{\We}$ as predicted by Eq.~(\ref{scaling}).
Such a dependence of $k$ on $\We$ explains the limited range of $\We$ for which such rebound is observed in our experiments. The two slanting lines bounding Region 1 for pancake bouncing on straight posts correspond to $w^2/(-b r_0\cos{\theta_Y})= 0.45$ and $1.5$ (Eq.~(\ref{scaling})). For almost all the experiments reported in the literature\cite{QuereReyssat2006,BartoloMoulinet2007,BhushanJung2008,WangMcCarthy2012,Tran2013}, this parameter takes values between $0.01$ and $0.144$, smaller than the threshold demonstrated in our work by at least one order of magnitude. On such surfaces, either the liquid penetration is insignificant (e.g., due to too narrow and/or too short posts) or the capillary energy stored can not be rectified into upward motion adequate to lift the drop (e.g., due to an unwanted Cassie-to-Wenzel transition \cite{Wenzel1936,Cassie1944,QuereLafuma2003,BartoloMoulinet2007, RobinVerho2012,ButtDeng2013}).
Region 2 shows that the introduction of tapered posts significantly widens the range of timescale and Weber number for pancake bouncing, way beyond Region 1. In this Region, the pancake bouncing can occur over a wider range of $k$ from 0.5 to 1.7 and $\We$ from 8.0 to 58.5. As emphasized above, for small $\We$ with moderate liquid penetration, the two timescales $t_{max}$ and $t_\uparrow$ are independent of $\We$. They become weakly dependent on $\We$ for relative large $\We$ due to the penetrated liquid hitting the base of the surface (Supplementary Movie\cite{SI}~5), but the emergence of pancake bouncing is rather insensitive to the post height as long as this is sufficient to allow for adequate capillary energy storage (Supplementary Fig.\cite{SI}~6). For much shorter posts, for example the tapered surface with a post height of  $0.3 \, \mathrm{mm}$, we did not observe the pancake bouncing due to insufficient energy storage.
\par
The novel pancake bouncing is also observed on a multi-layered, two-tier, superhydrophobic porous (MTS) surface (Supplementary Fig.\cite{SI}~7). The top layer of the MTS surface consists of a post array with post centre-to-centre spacing of $\sim 260 \, \mathrm{\mu m}$ and the underlying layers comprise a porous medium\cite{YarinLembach2010,ButtDeng2013} of pore size $\sim 200 \, \mathrm{\mu m}$, naturally forming a graded pathway for drop penetration and capillary emptying. The typical contact time of the drop with the MTS surface is $t_{contact} \sim $ 5.0\,ms (Supplementary Movie\cite{SI}~6) and the range of $\We$ is between 12 and 35 for pancake bouncing. These values are comparable to those on tapered surfaces. Taken together, our observations on tapered post surfaces and MTS surfaces demonstrate that the counter-intuitive pancake bouncing is a general and robust phenomenon. Moreover, there is enormous scope for designing structures to optimise pancake bouncing for multifunctional applications\cite{Yarin2006, Quere2008, PoulikakosJung2012, AizenbergMishchenko2010, Stone2012, JiangZheng2010}.

\section*{Methods}

\noindent \textbf{Preparation of tapered surface and straight post arrays.} The tapered surface with a size of $2.0 \times 2.0 \, \mathrm{cm^2}$ was created based on type 101 copper plate with a thickness of $3.18 \, \mathrm{mm}$ by combining a wire-cutting method and multiple chemical etching. Square posts arranged in a square lattice were first cut with a post centre-to-centre spacing of $200 \, \mathrm{\mu m}$. The post edge length and height are $100 \, \mathrm{\mu m}$ and $800 \, \mathrm{\mu m}$, respectively. Then the as-fabricated surface was ultrasonically cleaned in ethanol and deionized water for $10 \, \mathrm{min}$, respectively, followed by washing with diluted hydrochloric acid (1 M) for $10 \, \mathrm{s}$ to remove the native oxide layer. To achieve a tapered surface with post diameter of $20 \, \mathrm{\mu m}$ at the top, six cycles of etching were conducted. In each cycle, the as-fabricated surface was first immersed in a freshly mixed aqueous solution of $2.5 \, \mathrm{mol \, L^{-1}}$ sodium hydroxide and $0.1 \, \mathrm{mol \, L^{-1}}$ ammonium persulphate at room temperature for $\sim 60 \, \mathrm{min}$, followed by thorough rinsing with deionized water and drying in nitrogen stream. As a result of chemical etching, $\mathrm{CuO}$ nanoflowers with an average diameter $\sim 3.0 \, \mathrm{\mu m}$ were produced. Note that the etching rate at the top of the posts is roughly eight-fold of that at the bottom of the surface due to the formation of an etchant solution concentration gradient generated by the restricted spacing between the posts. To facilitate further etching, after each etching cycle the newly-etched surface was washed by diluted hydrochloric acid (1 M) for $10 \, \mathrm{s}$ to remove the oxide layer formed during the former etching cycle. Then another etching cycle was performed to sharpen the posts. In preparing the straight post arrays, only one etching cycle was conducted. All the surfaces were modified by silanization immersing in 1 mM n-hexane solution of trichloro(1H,1H,2H,2H-perfluorooctyl)silane for $\sim 60 \, \mathrm{min}$, followed by heat treatment at $\sim 150 \, \mathrm{{}^\circ C}$ in air for 1 hour to render surfaces superhydrophobic.
\\

\noindent \textbf{Preparation of multi-layered, two-tier, superhydrophobic porous (MTS) surface}. The MTS surface is fabricated on a copper foam with density $0.45 \, \mathrm{g/cm^3}$, porosity 94 \%, and thickness $0.16 \, \mathrm{cm}$. The nanostructure formation on the MTS surface and silanization were conducted using the same procedures described above.
\\

\noindent \textbf{Contact angle measurements.} The static contact angle on the as-prepared substrate was measured from sessile water drops with a ram\'{e}-hart M200 Standard Contact Angle Goniometer. Deionized water drops of $4.2 \, \mathrm{\mu L}$, at room temperature with $60 \%$ relative humidity, were deposited at a volume rate of $0.5 \, \mathrm{\mu L \, s^{-1}}$. The apparent, advancing ($\theta_a$) and receding contact angles ($\theta_r$) on the tapered surface with centre-to-centre spacing of $200 \, \mathrm{\mu m}$ are $165.6^{\circ} \pm 1.3^{\circ}$, $167.2^{\circ}\pm 1.1^{\circ}$ and $163.9^{\circ} \pm 1.4^{\circ}$, respectively. The apparent (equivalent to the intrinsic contact angle on tapered surface), advancing ($\theta_a$) and receding contact angle ($\theta_r$) on the surface with nanoflower structure alone are $160^{\circ}\pm 1.8^{\circ}$, $162.4^{\circ}\pm 2.8^{\circ}$, and $158.8^{\circ}\pm 1.7^{\circ}$, respectively. At least five individual measurements were performed on each substrate.
\\

\noindent \textbf{Impact experiments.} The whole experimental setup was placed in ambient environment, at room temperature with $60 \% $ relative humidity. Water drops of $\sim 13 \, \mathrm{\mu L}$ and $6 \, \mathrm{\mu L}$ (corresponding to radii $\sim 1.45 \, \mathrm{mm}$ and $1.10 \, \mathrm{mm}$, respectively) were generated from a fine needle equipped with a syringe pump (KD Scientific Inc.) from pre-determined heights. The dynamics of drop impact was recorded by a high speed camera (Fastcam SA4, Photron limited) at the frame rate of 10,000 fps with a shutter speed 1/93,000 sec, and the deformation of drops during impingement were recorded using ImageJ software (Version 1.46, National Institutes of Health, Bethesda, MD).

\bibliography{references_corr}

\vspace{30pt}
\small
\noindent \textbf{Acknowledgments}\\
We are grateful for support from the Hong Kong Early Career Scheme Grant (No. 125312), National Natural Science Foundation of China (No. 51276152), CityU9/CRF/13G, the National Basic Research Program of China (2012CB933302)  and Center of Super-Diamond and Advanced Films (COSDAF) at CityU to Z.W., ERC Advanced Grant, MiCE, to J. Y., and RGC Grant 603510 to T.Q.. Experimental assistance was provided by Yuan Liu and Lei Xu. The authors gratefully thank David Qu\'{e}r\'{e} for many useful discussions.

\noindent \textbf{Author contributions}\\
Z.W., Y.L., and L.M. conceived the research. Z.W. and J.Y. supervised the research. Y.L. designed and carried out the experiments. Y.L., L.M., and X.X. analysed the data. L.M., J.Y.,X.X., and T.Q. developed the model. Z.W., J.Y. and L.M. wrote the manuscript. Y.L. and L.M. contributed equally to this work.

\noindent \textbf{Additional information}\\
Supplementary information is available in the online version of the paper. Reprints and permissions information is available online at www.nature.com/reprints. Correspondence and requests for materials should be addressed to Z.W. (zuanwang@cityu.edu.hk) or J.Y. (j.yeomans1@physics.ox.ac.uk).


\noindent \textbf{Competing financial interests}\\
The authors declare no competing financial interests.

\newpage

\section*{Supplementary Information}

The Supplementary Information for this article is also available on the \href{http://www.nature.com/nphys/journal/vaop/ncurrent/abs/nphys2980.html}{Nature Physics website}, doi:10.1038/nphys2980, where the movies can be found.

\subsection*{Supplementary Movies}

\noindent \textbf{Supplementary Movie 1 $|$} The conventional complete rebound dynamics of water drop (radius $ r_0 \sim 1.45 \, \mathrm{mm}$) impacting on a tapered-post substrate with an impinging velocity $v_0 = 0.59 \, \mathrm{m/s}$, corresponding to  $\We=7.1$. The post centre-to-centre spacing is $200 \, \mathrm{\mu m}$. The frame rate set is $10,000 \, \mathrm{fps}$ with a shutter speed 1/93,000 sec and the movie playback speed is $29 \, \mathrm{fps}$.

\vspace{12pt}

\noindent \textbf{Supplementary Movie 2 $|$} The pancake bouncing dynamics of water drop (radius $r_0 \sim 1.45 \, \mathrm{mm}$) on the tapered-post substrate at $\We=14.1$. Here, the post centre-to-centre spacing is $200 \, \mathrm{\mu m}$. The frame rate set is $10,000 \, \mathrm{fps}$ with a shutter speed 1/93,000 sec and the movie playback speed is $29 \, \mathrm{fps}$.

\vspace{12pt}

\noindent \textbf{Supplementary Movie 3 $|$} Water drop (radius $r_0 \sim 1.45 \, \mathrm{mm}$) impact dynamics of on tilted  tapered-post (left) and nanostructured superhydrophobic surfaces (right). Here, the tilt angle is $30^{\circ}$ and $\We=31.2$. The post centre-to-centre spacing is $200 \, \mathrm{\mu m}$. The frame rate set is $10,000 \, \mathrm{fps}$ with a shutter speed 1/93,000 sec and the movie playback speed is $29 \, \mathrm{fps}$.

\vspace{12pt}

\noindent \textbf{Supplementary Movie 4 $|$} The impact dynamics of water drop (radius $r_0 \sim 1.45 \, \mathrm{mm}$) on the straight post substrate at $\We=7.9$. The post centre-to-centre spacing is $300 \, \mathrm{\mu m}$. The frame rate set is $10,000 \, \mathrm{fps}$ with a shutter speed 1/93,000 sec and the movie playback speed is $29 \, \mathrm{fps}$.

\vspace{12pt}

\noindent \textbf{Supplementary Movie 5 $|$} The impact dynamics of water drop (radius $r_0\sim 1.45 \, \mathrm{mm}$) on a tapered-post substrate at $\We=27.4$. The post centre-to-centre spacing is $200 \, \mathrm{\mu m}$. The frame rate set is $10,000 \, \mathrm{fps}$ with a shutter speed 1/93,000 sec and the movie playback speed is $29 \, \mathrm{fps}$.

\vspace{12pt}

\noindent \textbf{Supplementary Movie 6 $|$} The impact dynamics of water drop (radius $r_0 \sim 1.45 \, \mathrm{mm}$) on the MTS surface with an average pore size $200 \, \mathrm{\mu m}$ at $\We=19.7$. The frame rate set is $10,000 \, \mathrm{fps}$ with a shutter speed $1/93,000 \, \mathrm{sec}$ and the movie playback speed is $29 \, \mathrm{fps}$.

\newpage
\subsection*{Supplementary Figures}
\begin{figure}[h!] 
  \centering
  \includegraphics[width=0.5\textwidth]{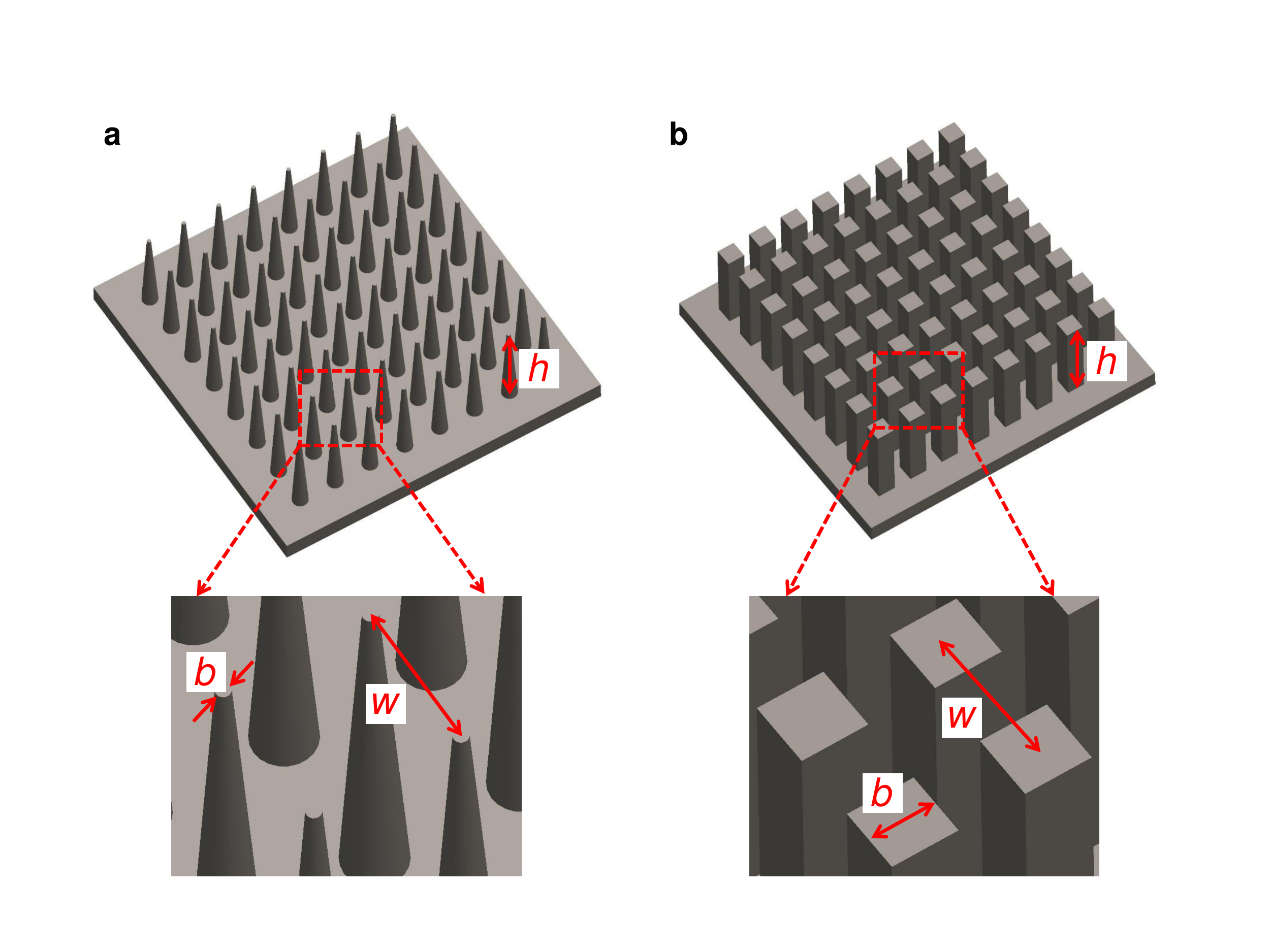}\\
	\caption{\textbf{Schematic drawing showing the surface architectures of tapered surface and straight post arrays.} Here, $b$  denotes the diameter of the post (the edge length for square post), $w$ is the center-to-center spacing, and $h$ is the post height.}
  \label{Fig:SIFig1}
\end{figure}
%
%
\begin{figure}[h!] 
  \centering
  \includegraphics[width=0.8\textwidth]{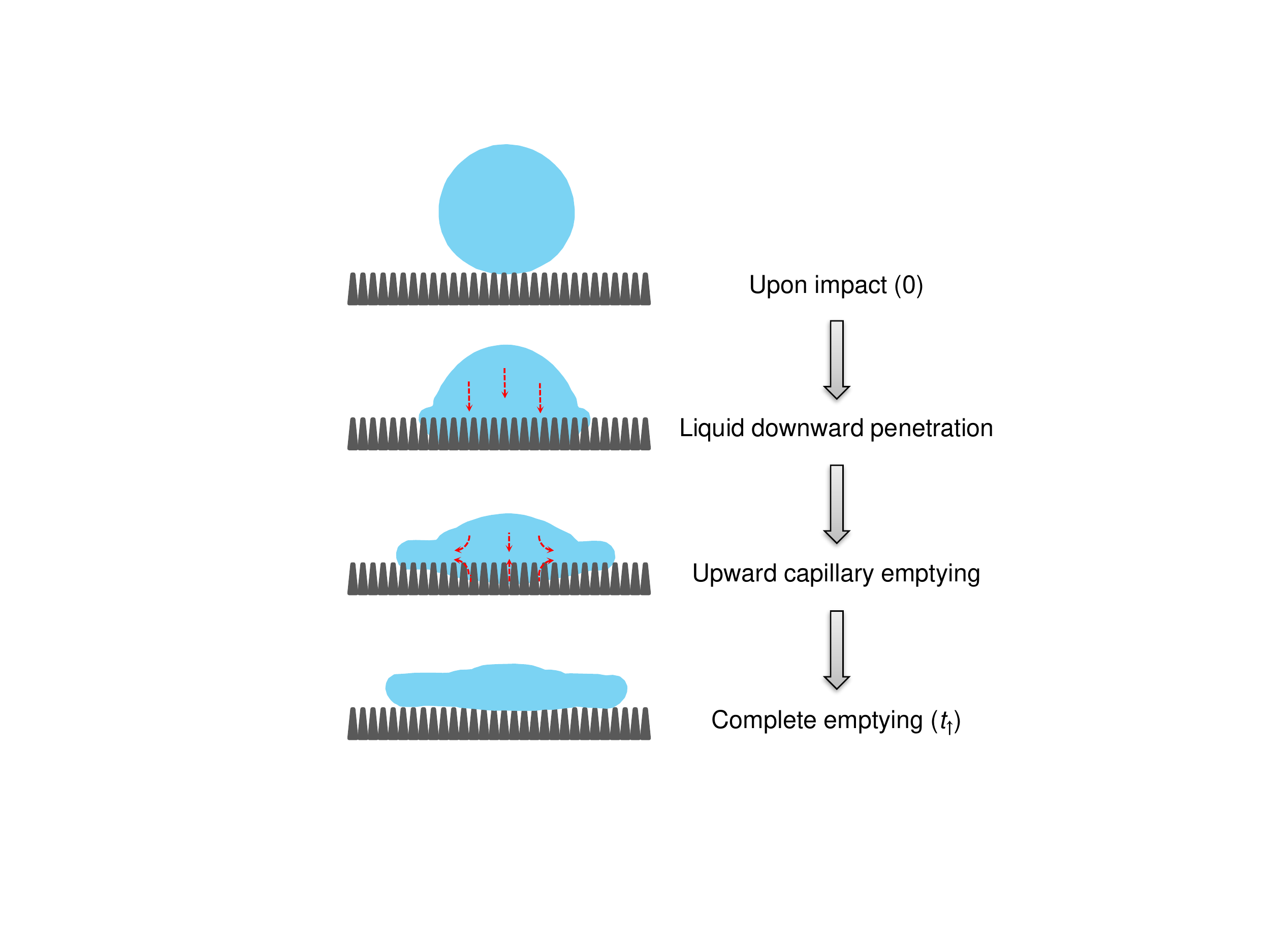}\\
	\caption{\textbf{Schematic drawing defining the timescale during drop impact.} $t_{\uparrow}$ is the time interval between the moment when the drop first touches the surface and when the substrate is completely emptied, during which fluid undergoes the downward penetration and upward capillary emptying processes.}
  \label{Fig:SIFig2}
\end{figure}
%
%
\begin{figure}[h!] 
  \centering
  \includegraphics[width=0.8\textwidth]{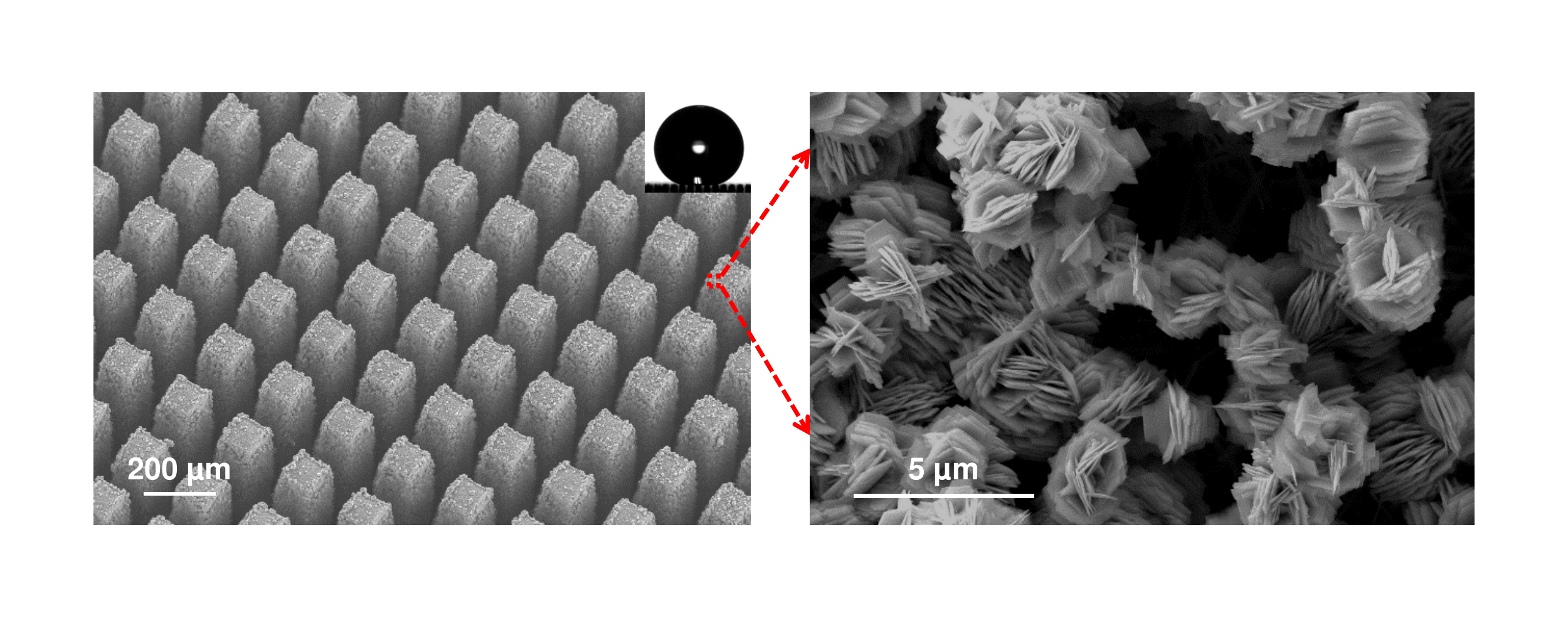}\\
	\caption{\textbf{SEM images of the straight post arrays.} The posts have a square cross section with a center-to-center post spacing of $200 \, \mathrm{\mu m}$, and they are uniformly covered by nanoflowers (right) to allow for a large intrinsic contact angle.
}
  \label{Fig:SIFig3}
\end{figure}
%
%
\begin{figure}[h!] 
  \centering
  \includegraphics[width=0.5\textwidth]{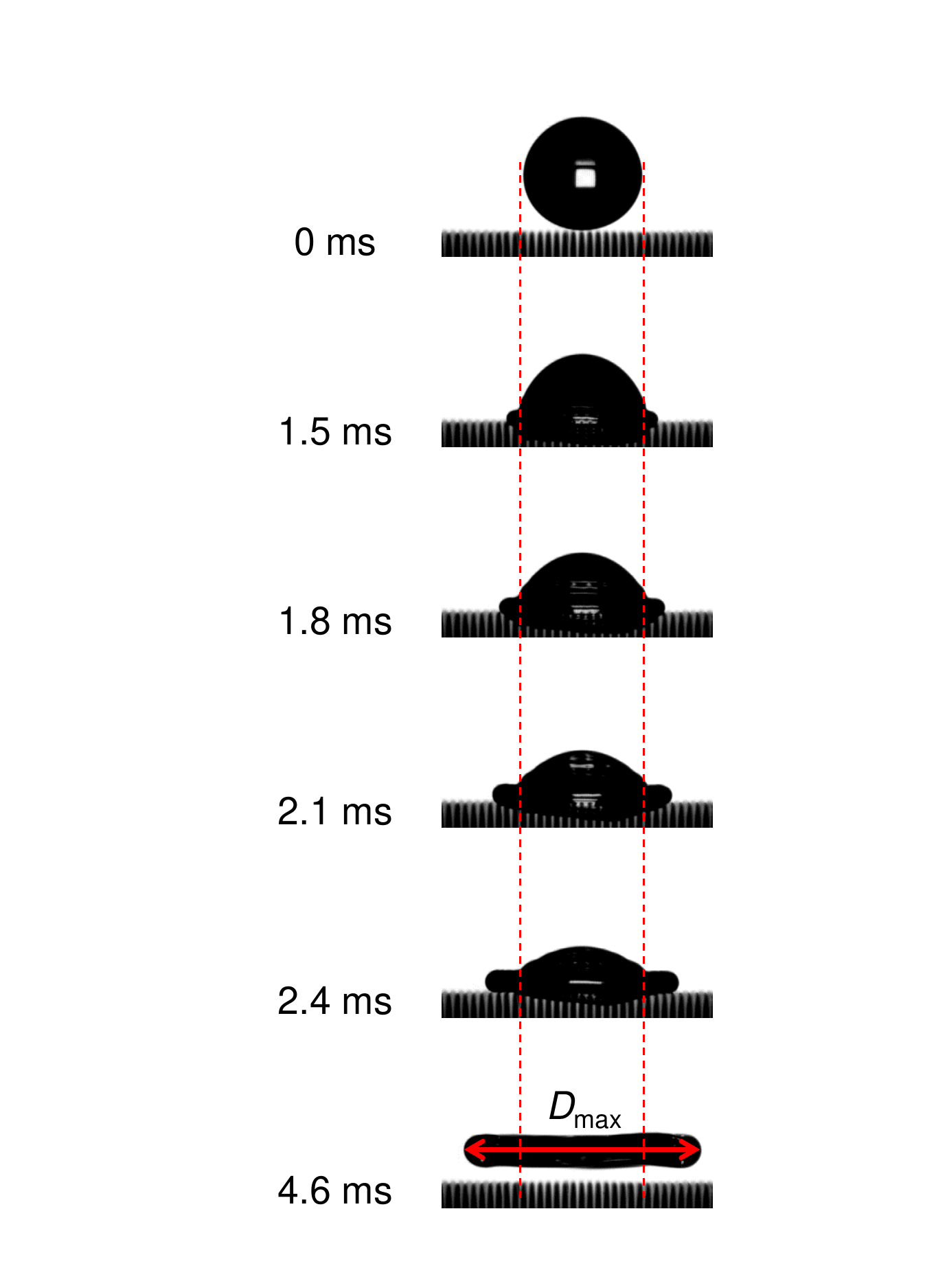}\\
	\caption{\textbf{Selected snapshots showing a drop impacting on the tapered surface with a post height of $\textbf{0.8} \,\mathrm{\textbf{mm}}$ under \textit{W\hspace{-0.25em}e}=14.1.} During the impact, the penetrating liquid, or equivalently the capillary energy, is primarily stored in a localized region. Careful inspection of the penetration dynamics reveals that the lateral extension of the localized region is approximately equivalent to the initial drop diameter ($2r_0$), which is much smaller than the maximal spreading diameter.}
  \label{Fig:SIFig4}
\end{figure}
%
%
\begin{figure}[h!] 
  \centering
  \includegraphics[width=0.8\textwidth]{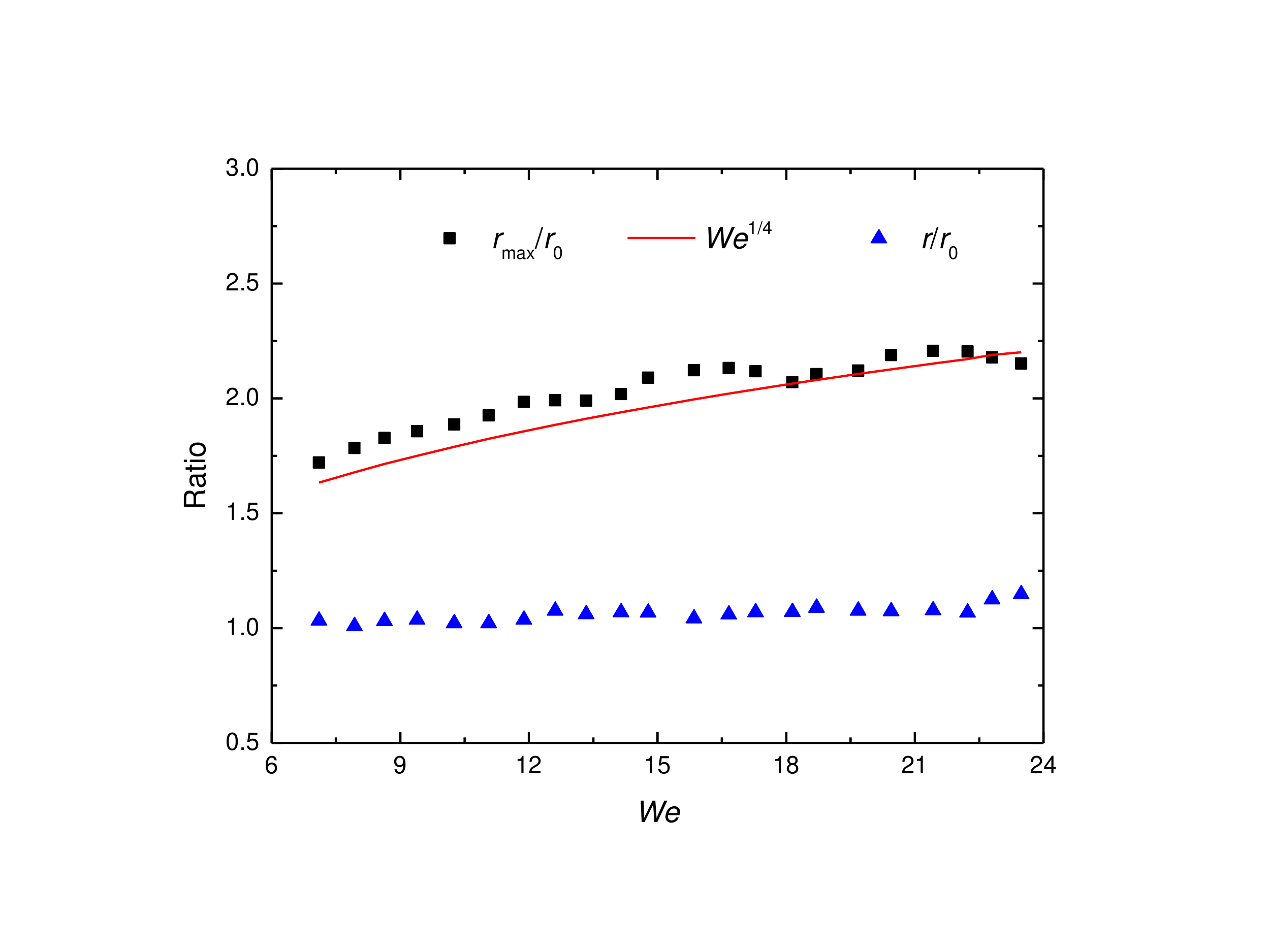}\\
	\caption{\textbf{The variation of the radius of the penetrating liquid within the substrate ($\textit{r}$) relative to the initial drop size ($\textit{r}_0$) as a function of \textit{W\hspace{-0.25em}e}.} It is apparent that the size of penetrating liquid is approximately equivalent to the initial drop size, which is independent of $\We$. Note that there is a large deviation between the measured $r$ and the maximal spreading radius (following $\sim r_0 \We^{1/4}$ , red solid line).
}
  \label{Fig:SIFig5}
\end{figure}
%
%
\begin{figure}[h!] 
  \centering
  \includegraphics[width=1.0\textwidth]{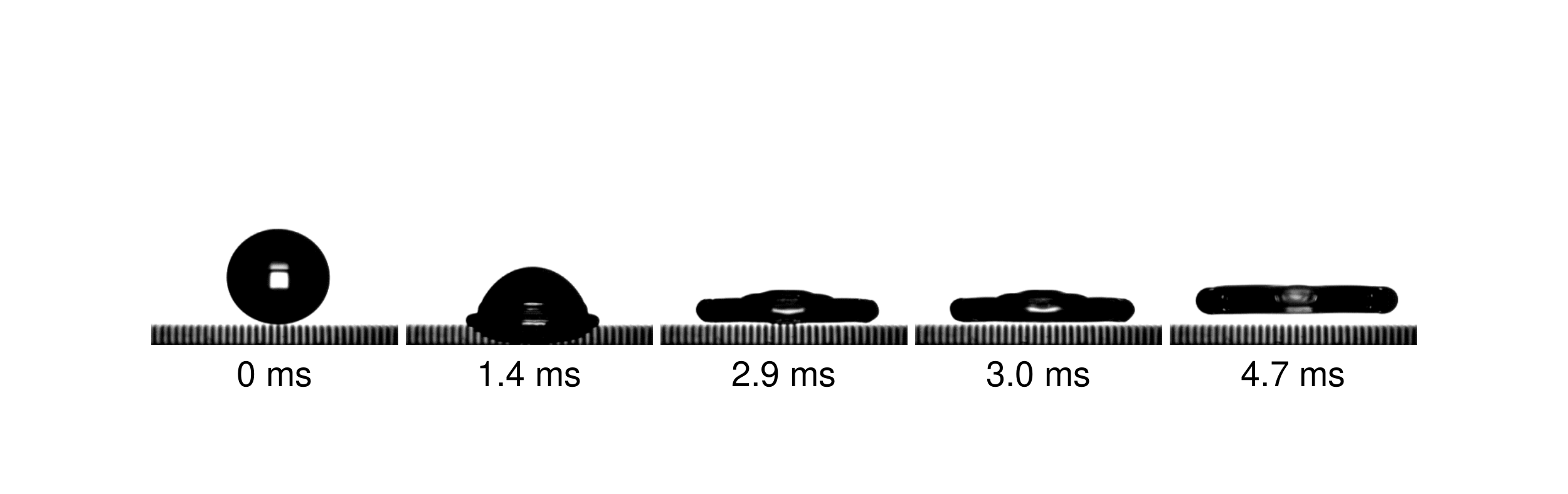}\\
	\caption{\textbf{Selected snapshots showing a drop impacting on tapered surfaces with post height of $\textbf{0.5} \,\mathrm{\textbf{mm}}$ under \textit{W\hspace{-0.25em}e}=14.1.} During the impact, the penetration liquid is localized in a region with a lateral extension roughly equivalent to its initial drop diameter. The penetrating liquid touches the base. The bouncing is very similar to that on the surface with $h = 0.8\, \mathrm{mm}$ (Fig. 1c in the main text), where the maximal drop penetration depth is smaller than the post height.
}
  \label{Fig:SIFig6}
\end{figure}
%
%
\begin{figure}[h!] 
  \centering
  \includegraphics[width=0.8\textwidth]{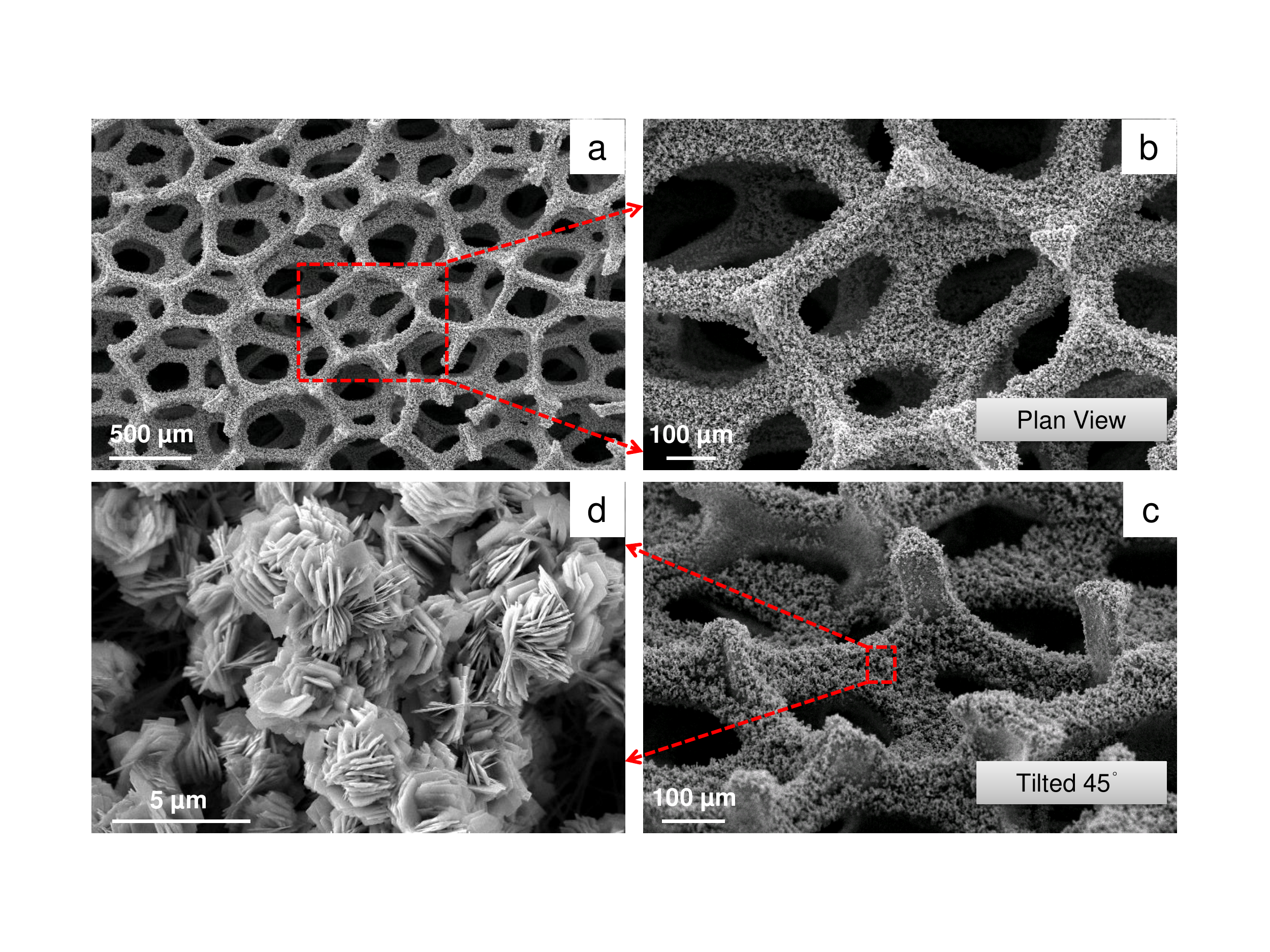}\\
	\caption{\textbf{SEM images of the multi-layered, two-tier, superhydrophobic porous (MTS) surface.} \textbf{a}, SEM images of the multi-layered, two-tier, superhydrophobic porous (MTS) surface. \textbf{b}, \textbf{c}, Top-down and $45^{\circ}$ side-view of the MTS substrate. The top layer of the substrate is a micro-post array in a hexagonal lattice arrangement which stands on a multi-layered porous medium. The post diameter, height and post centre-to-centre spacing of the top layer are $80  \, \mathrm{\mu m}$, $200  \, \mathrm{\mu m}$, and $260  \, \mathrm{\mu m}$, respectively. The average pore size in the porous media is $\sim 200 \, \mathrm{\mu m}$. \textbf{d}, SEM image showing the surface is uniformly covered by nanostructured flowers.
}
  \label{Fig:SIFig7}
\end{figure}

\end{document}